\documentclass[gi, manuscript]{copernicus}

\usepackage{bm}

\AtBeginDocument{\nolinenumbers}

\providecommand{\introductionname}{Introduction}
\providecommand{\conclusionname}{Conclusions}

\begin{document}

\title{Estimating sub-frame time differences in camera image sequences}

\Author[1]{Juha}{Vierinen}
\Author[1]{Pavithiran}{Sivasothy}
\Author[1]{Bj\"orn}{Gustavsson}

\affil[1]{University of Troms\o{}, The Arctic University of Troms\o{}, Hansine Hansens veg 18, 9019 Troms\o{}}

\correspondence{Juha Vierinen (jvi019@uit.no)}
\runningtitle{Technique for sub-frame camera calibration}
\runningauthor{J. Vierinen}

\received{}
\pubdiscuss{}
\revised{}
\accepted{}
\published{}
\firstpage{1}

\maketitle

\begin{abstract}
Some optical measurements require relative timing of intensity variations with accuracy much finer than the camera frame period. One motivating example is dynamic aurora, where different prompt emissions are expected to originate from different altitude regions and can therefore have millisecond-scale relative delays caused by finite energetic-electron velocities and other electron-transport effects. These delays are predicted to be a small fraction of the frame duration of typical auroral video cameras. We present a cross-spectral technique for estimating the relative delay between two time-varying optical intensity signals recorded by one or more image sensors. The method is validated with a calibration device that generates two pseudorandomly pulsed optical emissions with a known relative delay, recorded using a consumer smartphone camera. For the tested recordings, the method estimates relative delays between image-sensor regions with better than $50$~$\mu$s accuracy. Although developed for high-frame-rate auroral imaging, the technique has numerous other imaging applications, including camera timing calibration and measurements of time-varying optical signals. The single-camera tests demonstrate that the method can characterize sub-frame timing differences across an image sensor, such as those produced by rolling-shutter readout. The same analysis applies to separate cameras when they observe the same time-varying signal and are synchronized to a shared clock.
\end{abstract}
\section{Introduction}
\label{sec:intro}

For optical observations of aurora, synchronization requirements have traditionally been taken to be a fraction of the frame period and have been met using a GPS pulse-per-second signal. High-speed optical auroral observations have typically been limited to video frame rates, i.e. 25--50 frames per second (fps), for which a synchronization accuracy of about 5~ms is usually sufficient. Occasional observations at higher frame rates, for example by \citet{mcharg1998grl}, have predominantly been single-instrument observations for which explicit camera-to-camera synchronization is not required.

Recent time-dependent electron-transport calculations \citep{gustavsson2020jgr-tdetrp} predict that prompt emissions can exhibit time shifts during aurora that vary at frequencies above 5~Hz. Examples include flickering aurora \citep[e.g.][]{whiter2010jgr,sakanoi2004jgr}, auroral curls \citep{trondsen1998jgr,vogt1999jgr-auroral-curls,lanchester1997jgr}, and other dynamic small-scale auroras \citep[e.g.,][]{dahlgren2010jastp,semeter2006grl}. These time shifts arise from the finite velocity of precipitating primary electron pulses as they propagate downward in altitude, together with the energy-degradation time, which causes different prompt auroral emissions to occur with millisecond-scale offsets.

Resolving these small time shifts requires simultaneous observations in multiple auroral emissions with synchronization better than 1~ms, and preferably better than 100~$\mu$s. This is stricter than what has been achieved in previous ground-based multispectral high-speed imaging of flickering aurora by \citet{kataoka2011ground}, who sought to experimentally observe Alfv\'en wave acceleration. For camera synchronization, they used artificial flashes from an LED synchronized with a GPS signal, observed by two cameras simultaneously, to match the sampling time with an error of less than 10~ms.

In other fields, \citet{vibeck2015synchronization} obtained a synchronization error of 16.7~ms using a flash clock to synchronize two stereoscopic cameras in an autonomous-vehicle application. A video-stream synchronization method with precision better than 1~ms using rolling-shutter cameras was developed by \citet{smid2019rolling}. Their approach exploits the fact that each sensor row in a rolling-shutter camera starts its exposure with a small time offset relative to neighboring rows.

Reliable measurements of millisecond-scale time shifts in multi-wavelength auroral observations therefore require a new timing-estimation technique. The objective of this study is to present such a technique, demonstrate its validity, and characterize its achievable accuracy. The technique is based on phase detection of cross spectra between images of time-shifted signals, analogous to geodetic very long baseline interferometry methods for estimating relative time delays between radio receiver stations observing a point-like radio source \citep{shapiro19765}. To validate the technique in practice, we constructed a calibration device capable of producing optical emissions with known time shifts. Using image sequences recorded at 60~fps, we demonstrate detection of time shifts smaller than 1000~$\mu$s with uncertainty below 100~$\mu$s.

\section{Method}

We represent two time-varying intensity signals as $I_1(t)$ and $I_2(t)$. The signals are related by $I_2(t) = \gamma I_1(t-\tau)$, so the second signal is a scaled and time-shifted version of the first. Using the Fourier-transform convention below, the relationship between a time delay and the corresponding phase shift is
\begin{align}
I_1(\omega) &= \int_{-\infty}^{\infty} I_1(t) e^{-i\omega t}dt\\
I_2(\omega) &= \int_{-\infty}^{\infty} I_2(t) e^{-i\omega t}dt = \gamma e^{-i\omega \tau} \int_{-\infty}^{\infty} I_1(t') e^{-i\omega t'}dt = \gamma e^{-i\omega \tau} I_1(\omega).
\end{align}
The cross-spectrum of $I_1(t)$ and $I_2(t)$ therefore has a phase that depends on the time shift $\tau$. Assuming that $I_1(t)$ and $I_2(t)$ are random signals and that $\gamma$ is real and positive, the mean cross-spectrum is
\begin{align}
\mathrm{E}\{I_1(\omega) I_2^*(\omega)\} = \gamma e^{i\omega \tau} \mathrm{E}\{|I_1(\omega)|^2\}.    
\end{align}
The phase of the mean cross-spectrum is therefore linearly proportional to the time delay:
\begin{align}
\angle \mathrm{E} \{I_1(\omega) I_2^*(\omega)\} = \phi_{12}(\omega) = \omega \tau.    
\end{align}
Therefore, the phase of the mean cross-spectrum of two image pixels that measure the same intensity time series provides information about the relative time delay between the pixels.

In practice, the Fourier transforms of the discretized pixel-intensity signals are estimated using a discrete Fourier transform. These spectral estimates are corrupted by measurement noise. Provided that the time delay is constant over time, this noise can be reduced by averaging multiple independent measurements of the cross-spectrum.

The phase estimates from $N$ cross-spectral components can then be fit as a function of frequency using a linear least-squares model:
\begin{align}
\bm{m} &= \bm{A}\tau + \bm{\xi},\\
\begin{bmatrix}
\phi_{12}(\omega_1)\\
\phi_{12}(\omega_2)\\
\vdots \\
\phi_{12}(\omega_N)\\
\end{bmatrix} &= \begin{bmatrix}
\omega_1 \\
\omega_2 \\
\vdots \\
\omega_N 
\end{bmatrix}\tau +
\begin{bmatrix}
\xi_1 \\
\xi_2 \\
\vdots \\
\xi_N 
\end{bmatrix}
\end{align}
Here $\omega_n$ is the angular frequency, $\phi_{12}(\omega_n)$ is the measured cross-spectral phase at frequency $\omega_n$, and $\xi_n$ is the phase measurement error. For sufficiently large values of $|\tau|$, the measured phase must be unwrapped. This does not introduce additional measurement error if the phase measurement errors are sufficiently small, $\sigma \ll \pi$.

If the phase measurement errors are independent, normally distributed, zero-mean random variables, $\xi_n \sim \mathcal{N}(0,\sigma^2)$, the least-squares estimate is also the maximum-likelihood estimate:
\begin{equation}
\hat{\tau}_{\mathrm{ML}} = (\bm{A}^T \bm{A})^{-1}\bm{A}^T \bm{m}.
\end{equation}
Assuming that all phase errors are independent and identically distributed, the time-shift estimation error variance is
\begin{equation}
\mathrm{Var}\{\hat{\tau}_{\mathrm{ML}}\} = \sigma_{\tau}^2 = \sigma^2 (\bm{A}^T \bm{A})^{-1} .
\end{equation}
For this estimate to be useful, the light signal must contain nonzero spectral power over a sufficiently wide frequency range. A wider signal bandwidth improves the delay estimate, whereas signal power above the camera Nyquist frequency can alias into the observed spectrum and bias the phase slope.

\section{Measurements}

\begin{figure}
    \centering
    \includegraphics[width=0.38\textwidth]{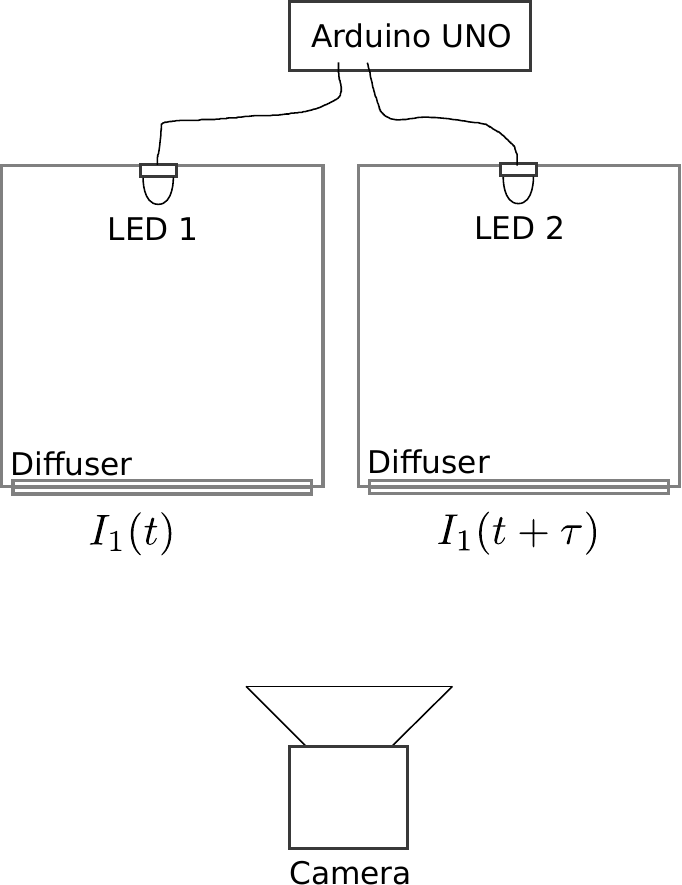}\hspace{0.2cm}\includegraphics[width=0.48\textwidth]{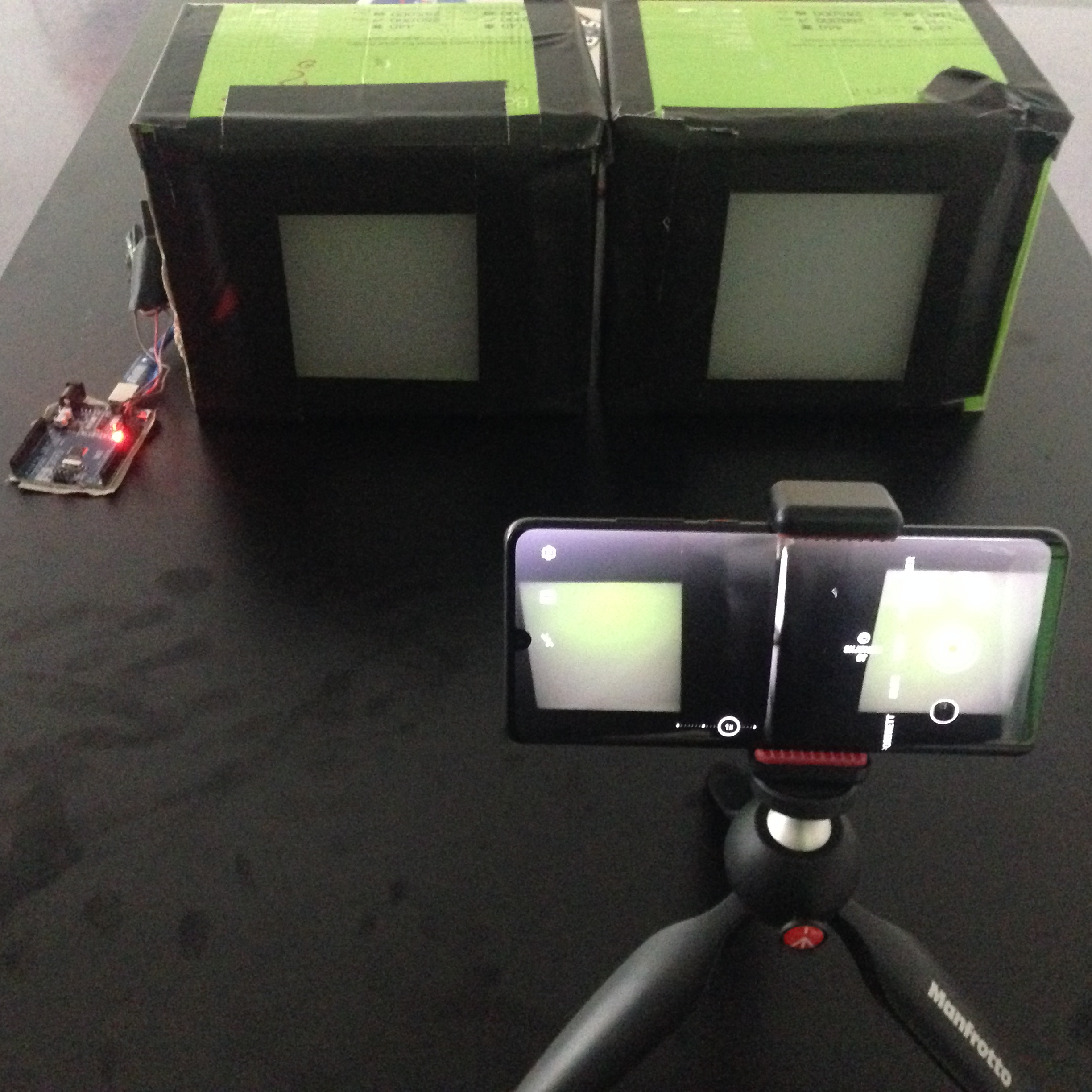}
    \caption{Left: diagram of the calibration device. Right: photograph of the calibration test setup.}
    \label{fig:device}
\end{figure}

\begin{figure}
    \centering
    \includegraphics[width=0.5\textwidth]{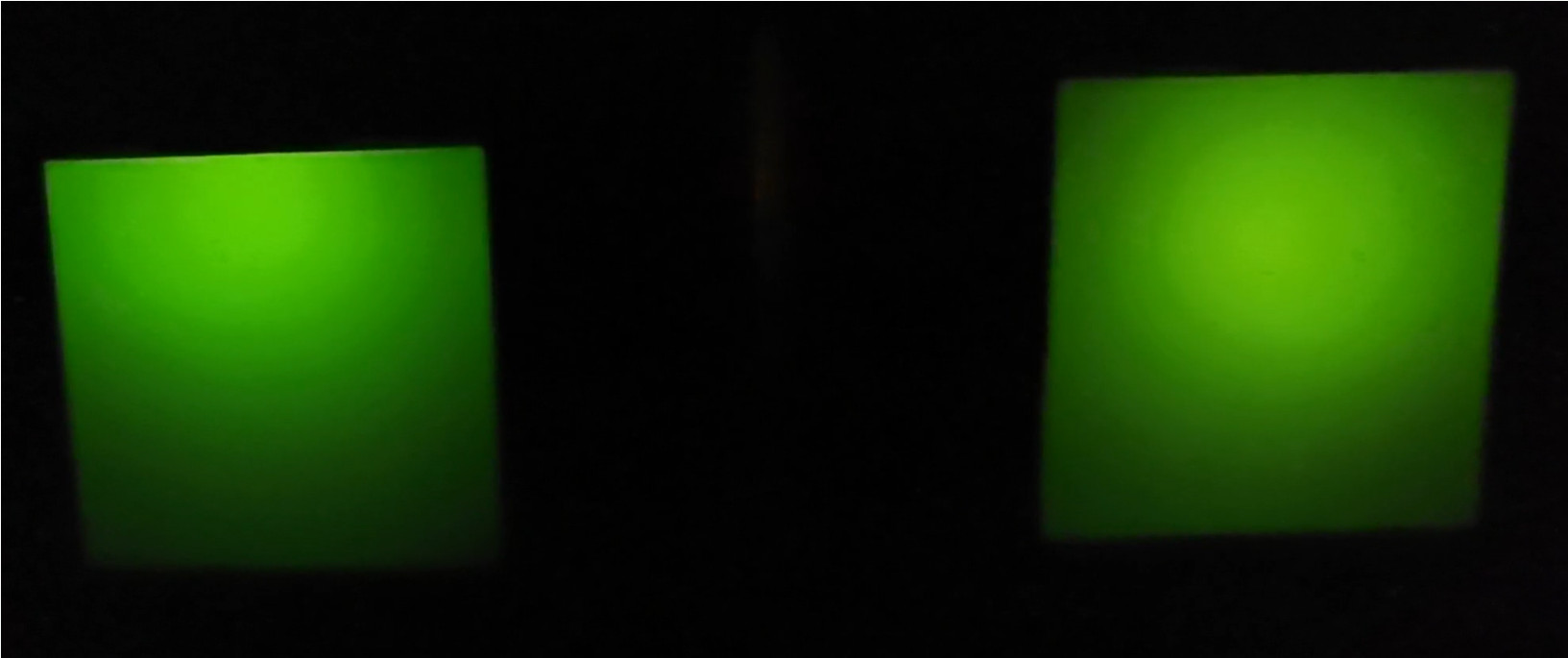}
    \caption{Single video frame of the calibration device. The region illuminated with intensity $I_1(t)$ is on the left, and the region illuminated with the time-shifted intensity $I_2(t)$ is on the right.}
    \label{fig:frame}
\end{figure}

To demonstrate and validate the technique for estimating relative time delays from camera time-series measurements, we constructed a calibrator that generates two regions of approximately spatially uniform light with time-varying pseudorandom intensity and a user-defined relative time shift. One source provides light with intensity $I_1(t)$, and the other provides light with intensity $\gamma I_1(t-\tau)$.

The light controller was implemented using an Arduino UNO microcontroller board, which controlled the on-off state of two light-emitting diodes (LEDs). The LEDs were located in two separate windowed boxes. The microcontroller was programmed to blink one LED on and off with a random pulse duration uniformly distributed between 10 and 200~ms. This randomness ensured that the signal contained spectral power over a broad frequency range while decreasing toward the edges of the spectrum. The second LED was programmed to reproduce the same temporal behavior as the first LED, but with a user-defined time shift $\tau$.

To create an approximately uniform illuminated region, a diffuser made from two layers of white methyl methacrylate resin (Biltema brand ``opal'' plexiglass) was placed on a cut-out side of each box. The boxes were otherwise sealed from light using black duct tape. A diagram and photograph of the calibrator are shown in Figure \ref{fig:device}. A single video frame of the device is shown in Figure \ref{fig:frame}.

The calibrator signal was recorded using the built-in smartphone camera on a Huawei P30 Pro. This phone records video with a resolution of $2336\times1080$ pixels at 60~fps. The recordings were made using a tripod and stored in H.264 MPEG-4 AVC format. Ten-minute recordings of pulsating light were made at several different time delays. Some tests were also repeated with the smartphone camera on an iPhone X at 30~fps to verify that the method is not specific to one sensor. The results were not significantly different with this camera, although the slightly noisier image sensor and lower frame rate produced somewhat larger estimation errors. We therefore omit the iPhone X measurements. All recordings were made in a dark room to avoid other time-synchronized light sources.

Although the technique and calibrator could be used to measure the time delay between two separate scientific cameras driven by a shared clock, it was technically simpler to demonstrate the method using a single camera and measuring relative time delays between different pixels on the sensor.

\begin{figure}
    \centering
    \includegraphics[width=0.69\textwidth]{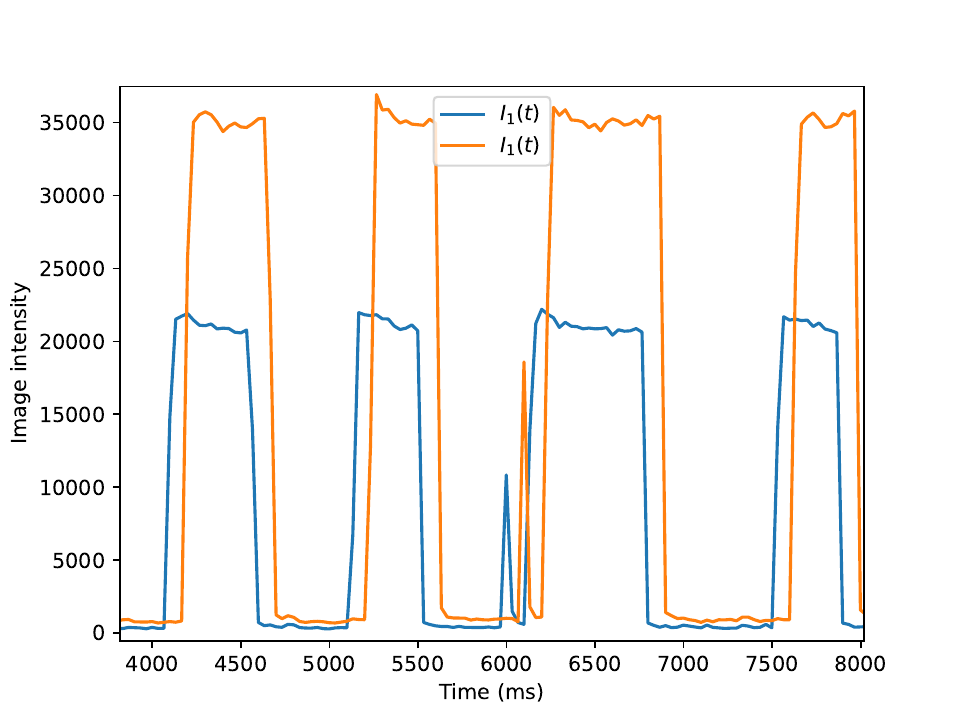}
    \caption{Image intensity $I_1(t)$ for one pixel on box 1 and image intensity $I_2(t)$ for one pixel on box 2. The time delay is set to $\tau=100$~ms.}
    \label{fig:signal}
\end{figure}

The two image intensity signals from one pixel on box 1 and one pixel on box 2 are shown in Figure \ref{fig:signal}. In this case, the device was configured to generate light pulses with a relative time delay of $\tau=100$~ms. Using a dual-channel oscilloscope, the error in the relative time delay between the two leading edges of the signals was determined to be less than 1~$\mu$s.

\section{Results}

This section presents measurements obtained with the calibration device. We show examples of cross-spectral power, cross-spectral phase, and least-squares fits of time delay to the phase measurements. We also show how the measurement error depends on the length of the video recording. Finally, we demonstrate relative-delay measurements for different pixels on an image sensor, allowing characterization of the sensor rolling shutter.

\subsection{Cross-spectral power and phase}

\begin{figure}
    \centering
    \includegraphics[width=\textwidth]{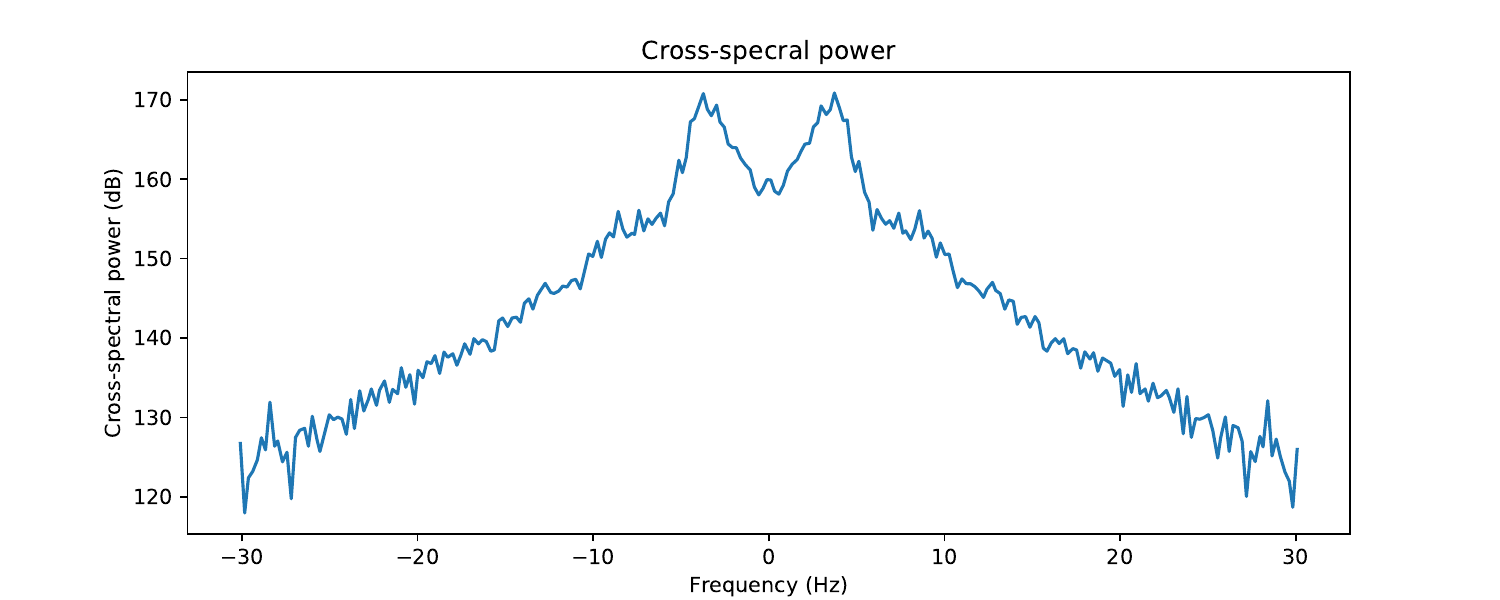}
    \caption{Cross-spectral power $10\log_{10}(|I_1(\omega)I_2^*(\omega)|)$ corresponding to a 2.8~min recording of a calibration signal.}
    \label{fig:xc}
\end{figure}

\begin{figure}
    \centering
    \includegraphics[width=0.8\linewidth]{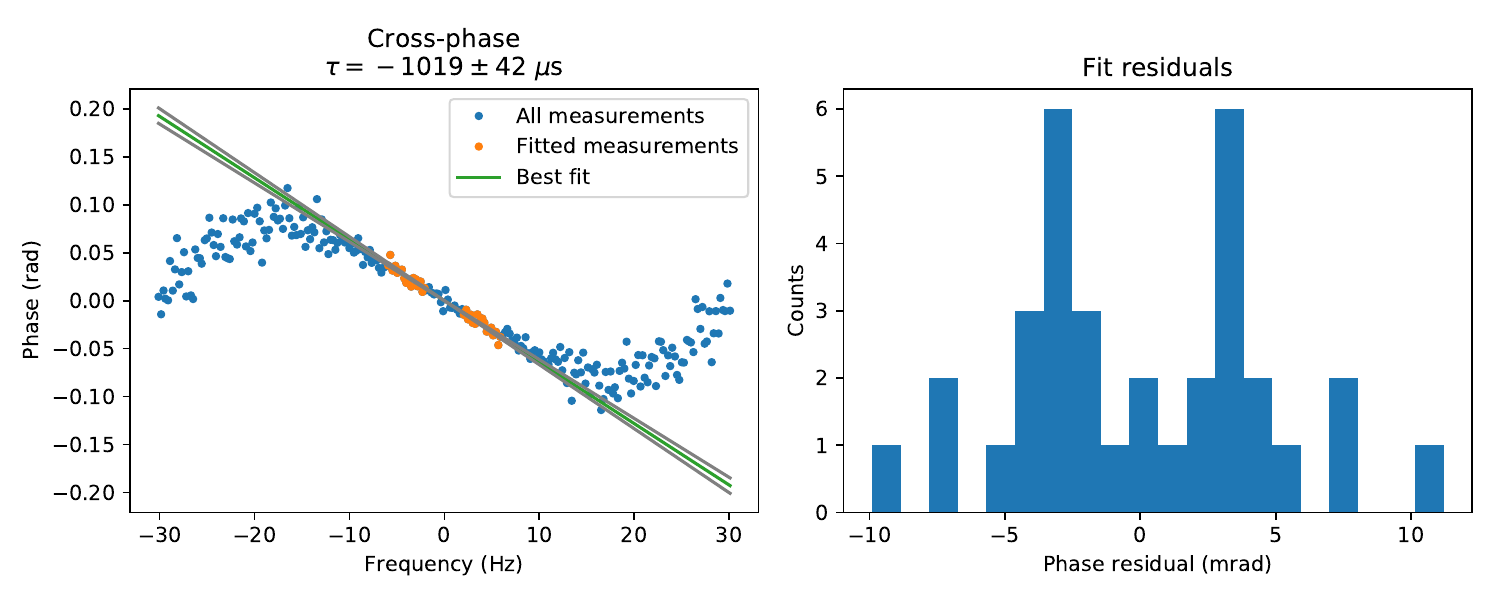}\par\medskip
    \includegraphics[width=0.8\linewidth]{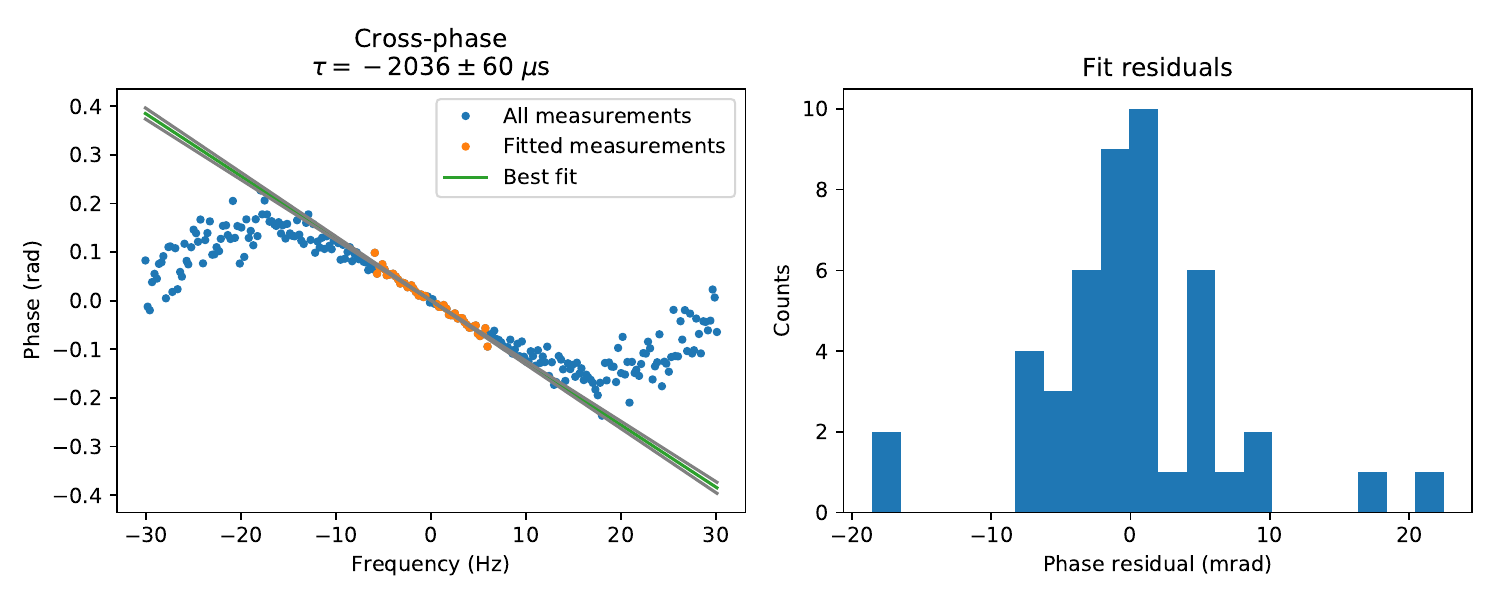}\par\medskip
    \includegraphics[width=0.8\linewidth]{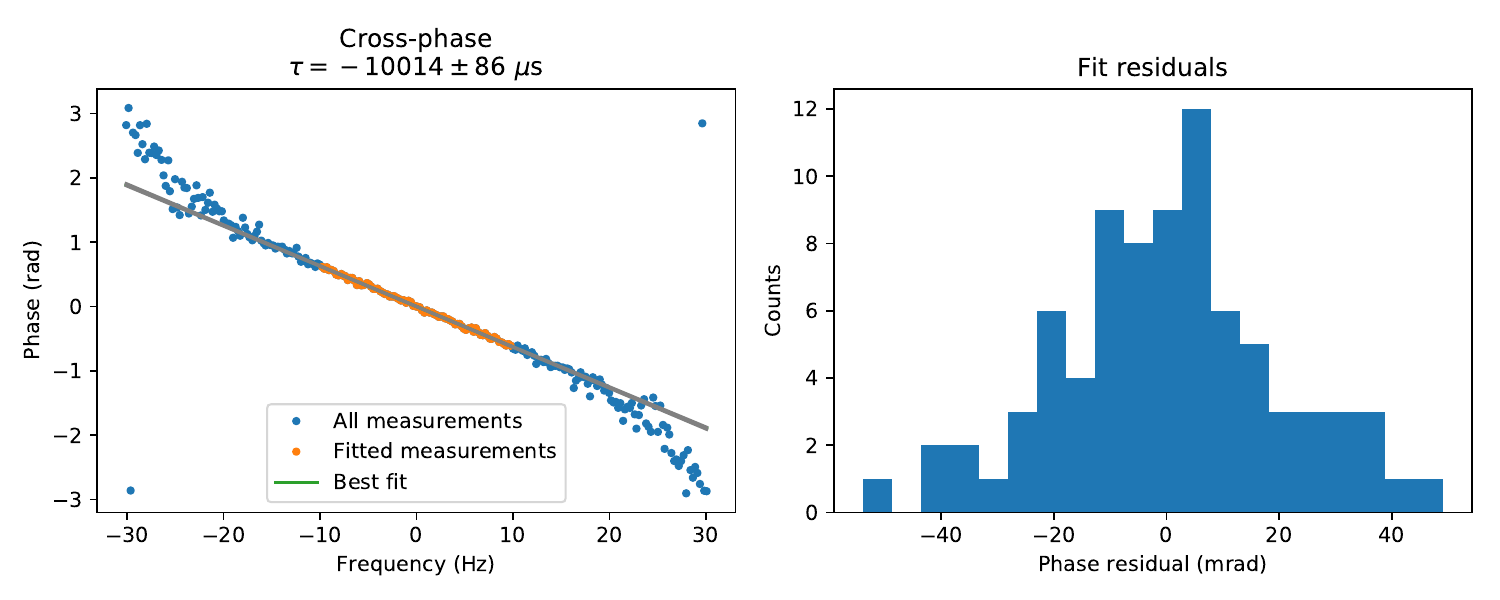}
    \caption{Cross-spectral phase measurements and corresponding maximum-likelihood linear fits are shown on the left. Fit residuals are shown on the right in milliradians. Depending on the measurement, only the linear part of the phase slope is used for the fit. Top: 1.05~ms, middle: 2.05~ms, and bottom: 10.05~ms time delay. All fits are within 50~$\mu$s of the true delay set by the calibrator.}
    \label{fig:xphase-fits-n-residuals}
\end{figure}

Figure \ref{fig:xc} shows the estimated cross-spectral power for a 2.8~min recording of signals $I_1(t)$ and $I_2(t)=\gamma I_1(t-\tau)$ with $\tau=10$~ms. Power is shown in decibel units. In this case, image pixels centered on the two calibration sources and located on the same sensor row are used to estimate the cross-spectrum. The spectral power is concentrated mostly at low frequencies and gradually decreases toward the edges of the spectrum. This is intentional, because aliasing can produce unexpected phase behavior.
Cross-spectral phase measurements for several 2.8~min recordings of the calibrator source are shown in Figure \ref{fig:xphase-fits-n-residuals}. For these measurements, we averaged 40 neighboring cross-spectral components obtained from discrete Fourier transforms of $I_1(t)$ and $I_2(t)$. The signals were selected as intensity values of single pixels, and three time delays are shown: 1.05, 2.05, and 10.05~ms. The central part of the cross-spectral phase follows a linear slope, $\phi_{12}(\omega)=\omega \tau$, which provides the delay estimate. The phase near the spectral edges does not follow a linear slope. The reason is uncertain, but possible causes include aliased signals, low spectral power at higher frequencies, and video-compression artifacts. We therefore use only the central linear part of the cross-spectral phase when fitting for time delay. All fits are within 50~$\mu$s of the true delay set by the calibrator.

\subsection{Measurement uncertainty}

To characterize the error behavior as a function of measurement duration, we conducted a series of measurements with different analysis time spans and estimated the standard deviation of the time-delay estimate. We used a time delay of $\tau=1.05$~ms for all tests. The phase measurement error variance was estimated from the maximum-likelihood fit residuals. All spectral components with frequencies between 0 and 10~Hz were used to fit the time delay; negative spectral components were not used, to reduce correlation between errors.

The measurement durations, number of frames, estimated standard deviations, estimated time delays, and errors relative to the true delay are shown in Table \ref{tab:errors}. As expected, the estimated standard deviation decreases as the measurement duration increases. The absolute error also generally decreases with duration. The observed errors are within a few estimated standard deviations of the true delay, suggesting that the simple independent equal-variance phase-error model captures the main scaling but may slightly underestimate the uncertainty for some recordings. This could be caused by lower-quality cross-spectral phase measurements at higher frequencies, aliasing, or video-compression artifacts.

\begin{table}
    \caption{Measurements of time shift for different measurement durations. All measurements use a $\tau=1050$~$\mu$s delay setting. The measurement duration is $T$, the number of analyzed image frames is $N_f$, the standard deviation of the time-shift estimate is $\sigma_{\tau}$, the estimated time shift is $\hat{\tau}_{\mathrm{ML}}$, and the error is $\hat{\tau}_{\mathrm{ML}}-\tau$.}
    \centering
    \begin{tabular}{c|c|c|c|c}
    \hline
        $T~(s)$ &  $N_f$ & $\sigma_{\tau}$~~($\mu$s) & $\hat{\tau}_{\mathrm{ML}}$~($\mu$s) & $\hat{\tau}_{\mathrm{ML}} - \tau$~~($\mu$s)\\
        \hline
        \hline
        5 & 300  &  279  & 1255 & 205\\
        17 & 1000 & 72 & 962 & -88 \\
        50 & 3000 & 40 & 955 & -95 \\
        167 & 10000 & 26 & 1018 & -32\\
         \hline
    \end{tabular}
    \label{tab:errors}
\end{table}
Some artifacts of video compression are expected. To avoid overly optimistic results, we do not present results for neighboring pixels, because their intensities can be highly correlated by the video compression algorithm. However, it is possible that some light intensity from one calibrator region leaks into the other region through video compression, which would cause the measured delay to be underestimated slightly.

\subsection{Sensor characterization}

\begin{figure}
    \centering
    \includegraphics[width=1\textwidth]{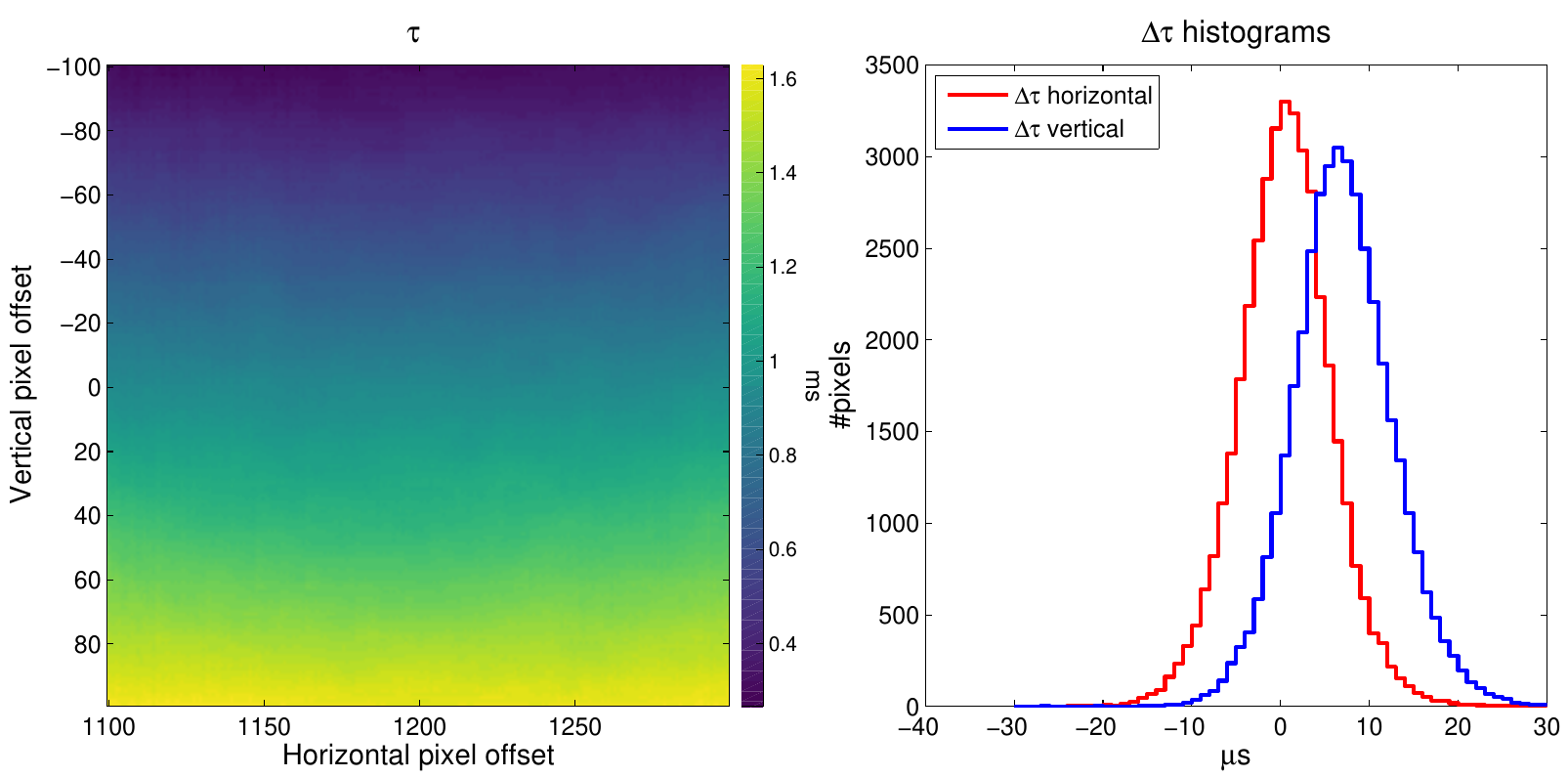}
    \caption{Time delays, $\tau$, as a function of pixel offset in milliseconds when comparing one pixel on light source 1 with pixels on light source 2 are shown in the left panel. The linear slope is caused by the rolling shutter of the smartphone camera used for this test. With zero vertical pixel offset, the time delays are close to the 1~ms delay generated by the calibrator. The distribution of time-delay differences between adjacent pixels is shown in the right panel. The change in time shift between adjacent pixels in the same row is small, while the time-shift change from one row to the next is, on average,
    6.58~$\mu$s.}
    \label{fig:rolling_shutter}
\end{figure}

To demonstrate the variation of time delay as a function of image pixel, we estimated delays for a $200\times200$ pixel region centered on the second light source. The time delays for these pixels were measured relative to a single image pixel on the first light source. For each pixel, a time delay was estimated by fitting the cross-spectral phase. The calibration device was configured to produce a time delay of $\tau=1$~ms between the two light sources. We used $10^4$ image frames, corresponding to a 167~s time span, to estimate the delays.

The variation of estimated time delay as a function of image pixel is shown in Figure \ref{fig:rolling_shutter}. The main visible feature is the rolling-shutter effect. Time delays on each row are relatively similar to one another, but the delay gradually increases as a function of image row. With zero vertical pixel offset, the delays are close to the 1~ms delay configured on the calibrator. The distribution of time-delay differences between adjacent pixels is shown in the right panel of the figure. The change in time shift between adjacent pixels in the same row is small, while the time-shift change from one row to the next is, on average, 6.58~$\mu$s. This is a plausible row-readout time. The histograms of adjacent-pixel time-delay differences suggest that the standard deviation of the measurements is less than $10$~$\mu$s, but this measurement does not by itself determine absolute accuracy. In the previous section, we found an absolute accuracy of about 30~$\mu$s with $10^4$ image frames. This discrepancy could be due, for example, to sample-clock frequency errors, but the most likely explanation is that 30~$\mu$s is a conservative estimate of the uncertainty.

\section{Conclusions}

The results show that the proposed technique can estimate sub-frame relative time delays between image pixels recorded by a digital video camera. The technique can be used to characterize image-sensor timing and to measure sub-frame time shifts in spectrally broad image time series recorded with one or more cameras.

The camera-timing calibration device is inexpensive and simple to build. By using only one of the light sources, it is possible to generate a zero-delay signal that can be used for inter-camera and inter-sensor time calibration. The dual-source signal generator allowed us to verify that a known time delay can be recovered accurately using the proposed technique.

With sufficiently long recordings of calibration signals with known time shifts, it is possible to estimate the time delay between image pixels with accuracy better than 50~$\mu$s. This is much shorter than the duration of an image frame. The accuracy is also sufficient to measure the rolling-shutter timing of image pixels read by a CMOS or other scanned-readout camera. Characterization of a region of pixels on the image sensor suggests that the standard deviation of relative time-delay errors can be below 10~$\mu$s. These values are specific to the camera, video compression algorithm, calibrator setup, and measurement duration used here.

The measurement accuracy is sufficient for studies of electron-transport effects in dynamic aurora, where the relative delay between prompt emissions from different altitudes during flickering aurora is expected to be several milliseconds. This study shows that it is feasible to measure such delays to within 100~$\mu$s for events lasting less than 20~s, provided that the auroral emission has sufficient temporal bandwidth, here approximately 10~Hz. Application of this technique to dynamic aurora and time-dependent electron transport is a topic for future work.

The technique should also be useful in other digital-imaging applications that require video-camera pixel timing. One possible extension would be to estimate the relative clock frequencies of two cameras when the clocks are not locked, in addition to estimating the relative time delays of the image-sensor readouts.

\codedataavailability{The analysis software and calibrator microcontroller code are available at \url{https://github.com/jvierine/image_xc_delay}.}

\authorcontribution{J. Vierinen contributed to this article by developing the method, software, and hardware. P. Sivasothy contributed by participating in the development of the hardware and software and by performing the measurements. B. Gustavsson contributed by developing the method and providing the scientific motivation. All authors contributed to preparing the manuscript.}

\competinginterests{No competing interests.}

\bibliographystyle{copernicus}
\bibliography{references.bib}

\end{document}